\documentclass[pre,twocolumn,superscriptaddress]{revtex4-1}  

\usepackage[paperwidth=210mm,paperheight=297mm,centering,hmargin=2cm,vmargin=2.5cm]{geometry}

\usepackage[pdftex]{graphicx}
\usepackage{amsmath,amsfonts,amssymb}
\usepackage{morefloats}
\usepackage{slashbox}
\usepackage{natbib}

\usepackage{hyperref}
\usepackage{xcolor} 

\definecolor{bluemoi}{rgb}{0.25,0.50 ,0.75} 

\hypersetup{
  bookmarksopen=true,
  pdftitle=" ",
  pdfauthor=" ", 
  pdfsubject=" ", 
  pdftoolbar=true, 
  pdfmenubar=true, 
  pdfhighlight=/O, 
  colorlinks=true, 
  pdfpagemode=UseNone, 
  pdfpagelayout=SinglePage, 
  pdffitwindow=true, 
  linkcolor=bluemoi, 
  citecolor=bluemoi, 
  urlcolor=bluemoi 
}

\makeatletter
\renewcommand{\figurename}{\sf \textbf{Figure}}
\renewcommand{\thefigure}{\arabic{figure}}
\renewcommand{\fnum@figure}{\sf\textbf{\figurename}~\textbf{\thefigure}}
\renewcommand{\tablename}{\sf\textbf{Table}}
\renewcommand{\thetable}{\arabic{table}}
\renewcommand{\fnum@table}{\sf\textbf{\tablename}~\textbf{\thetable}}
\makeatother

\begin{document}

\title{Comparing and modeling land use organization in cities}

\author{Maxime Lenormand}\affiliation{Instituto de F\'isica Interdisciplinar y Sistemas Complejos IFISC (CSIC-UIB), Campus UIB, 07122 Palma de Mallorca, Spain}
\author{Miguel Picornell}\affiliation{Nommon Solutions and Technologies, calle Ca\~nas 8, 28043 Madrid, Spain}	
\author{Oliva G. Cant\'u-Ros}\affiliation{Nommon Solutions and Technologies, calle Ca\~nas 8, 28043 Madrid, Spain}
\author{Thomas Louail}\affiliation{Institut de Physique Th\'{e}orique, CEA-CNRS (URA 2306), F-91191, Gif-sur-Yvette, France}\affiliation{G\'eographie-Cit\'es, CNRS-Paris 1-Paris 7 (UMR 8504), 13 rue du four, FR-75006 Paris, France}
\author{Ricardo Herranz}\affiliation{Nommon Solutions and Technologies, calle Ca\~nas 8, 28043 Madrid, Spain}
\author{Marc Barthelemy}\affiliation{Institut de Physique Th\'{e}orique, CEA-CNRS (URA 2306), F-91191, Gif-sur-Yvette, France}\affiliation{Centre d'Analyse et de Math\'ematique Sociales, EHESS-CNRS (UMR 8557), 190-198 avenue de France, FR-75013 Paris, France}
\author{Enrique Fr\'{i}as-Mart\'{i}nez}\affiliation{Telef\'{o}nica Research, 28050 Madrid, Spain}
\author{Maxi San Miguel}\affiliation{Instituto de F\'isica Interdisciplinar y Sistemas Complejos IFISC (CSIC-UIB), Campus UIB, 07122 Palma de Mallorca, Spain}
\author{Jos\'e J. Ramasco}\affiliation{Instituto de F\'isica Interdisciplinar y Sistemas Complejos IFISC (CSIC-UIB), Campus UIB, 07122 Palma de Mallorca, Spain}

\begin{abstract} 
The advent of geolocated ICT technologies opens the possibility of exploring how people use space in cities, bringing an important new tool for urban scientists and planners, especially for regions where data is scarce or not available. Here we apply a functional network approach to determine land use patterns from mobile phone records. The versatility of the method allows us to run a systematic comparison between Spanish cities of various sizes. The method detects four major land use types that correspond to different temporal patterns. The proportion of these types, their spatial organization and scaling show a strong similarity between all cities that breaks down at a very local scale, where land use mixing is specific to each urban area. Finally, we introduce a model inspired by Schelling's segregation, able to explain and reproduce these results with simple interaction rules between different land uses.  
\end{abstract}

\maketitle

\section*{INTRODUCTION}

Land use patterns appear as a natural result of citizens' and planners' interaction with the urban space. However, in a feedback loop, they also play a major role in the experience that residents and visitors have of a city \cite{Humphries2012}. Land use patterns have an effect on the liveability of neighborhoods and even on the health of the local residents \cite{Sallis2009}. On the other hand, land use and transportation display a well-established relation \cite{Badoe2000,Cervero2002,Krygsman2004,Tsai2005}. Transport demand depends on the location of residence and business areas, while the presence of new transport lines or facilities such as metro stations can substantially modify the land use mixing in a given area of the city. These ideas lie behind the development of the so-called Land Use Transport Interaction (LUTI) models \cite{Cervero1997,Waddell2007}, which are commonly employed in transport planning around the globe \cite{Bartholomew2007}.

An important issue regarding land use refers to the methods employed to estimate it. City Hall registers, surveys or satellite images have been used in the past to this end \cite{cadastro,Donnay2001,Yang2002,Geurs2004,Wu2006,Frias2012,Hu2013}. The emergence of geo-located ICT technologies introduces extra capabilities to directly measure the use that citizens make of each urban space. The information is exhaustive in terms of spatial and temporal resolution, allowing for the detection of concentrations of people second by second along days, weeks and months. Information from mobile phone call records \cite{Reades2007,Gonzalez2008,Reades2009,Soto2011,Toole2012,Noulas2013,Pei2013,Grauwin2014,Frias2014,Frias2014b,Lenormand2014,Louail2014,Deville2014,Louail2015}, geolocated tweets \cite{Frias2012,Goncalves2014,Hawelka2014,Lenormand2014b,Lenormand2015b}, credit card use \cite{Lenormand2015a} or FourSquare \cite{Noulas2013} has been considered in the literature. Different data sources have been compared, finding a consistent agreement among the estimations on human concentrations and mobility obtained from different ICT data \cite{Lenormand2014}, as well as between ICT data and more traditional techniques \cite{Toole2012,Noulas2013,Pei2013,Grauwin2014,Lenormand2014,Deville2014,Tizzoni2014}.

Such wealth of information together with the ability to process massive data brought by the Internet era allows the systematic comparison of features across cities. This analysis can lead to the discovery and confirmation of properties that have been hypothesized to be common to all cities, and also to  laws providing insights into the way a property scales with city size. Some examples of these properties include number of patents filed, unemployment rates, GDP per capita, business diversity, consumption of resources, length of road networks, or even crime density \cite{Bettencourt2007,Batty2008,Bettencourt2010,Bettencourt2013,Batty2013,Arcaute2015,Alves2013}. The finding of these laws raises the hope of the existence of a coherent framework for city science \cite{Makse1998,Batty2008,Bettencourt2010,Batty2013,Arcaute2015,Louf2013,Louf2014}. 

In this work, we explore land use patterns in the five most populous urban areas of Spain. Land use information is obtained from mobile phone records using a new framework based on network theory and systematic comparisons of land use distribution across the five cities are performed at different scales. Our results reveal common features in the land use types' spatial distributions, which can be understood with a model introduced also here. The similarities break down when the land use type mixing is studied at very short spatial scales, exposing patterns characteristic to each city.

\begin{figure}
\begin{center}
\includegraphics[width=8cm]{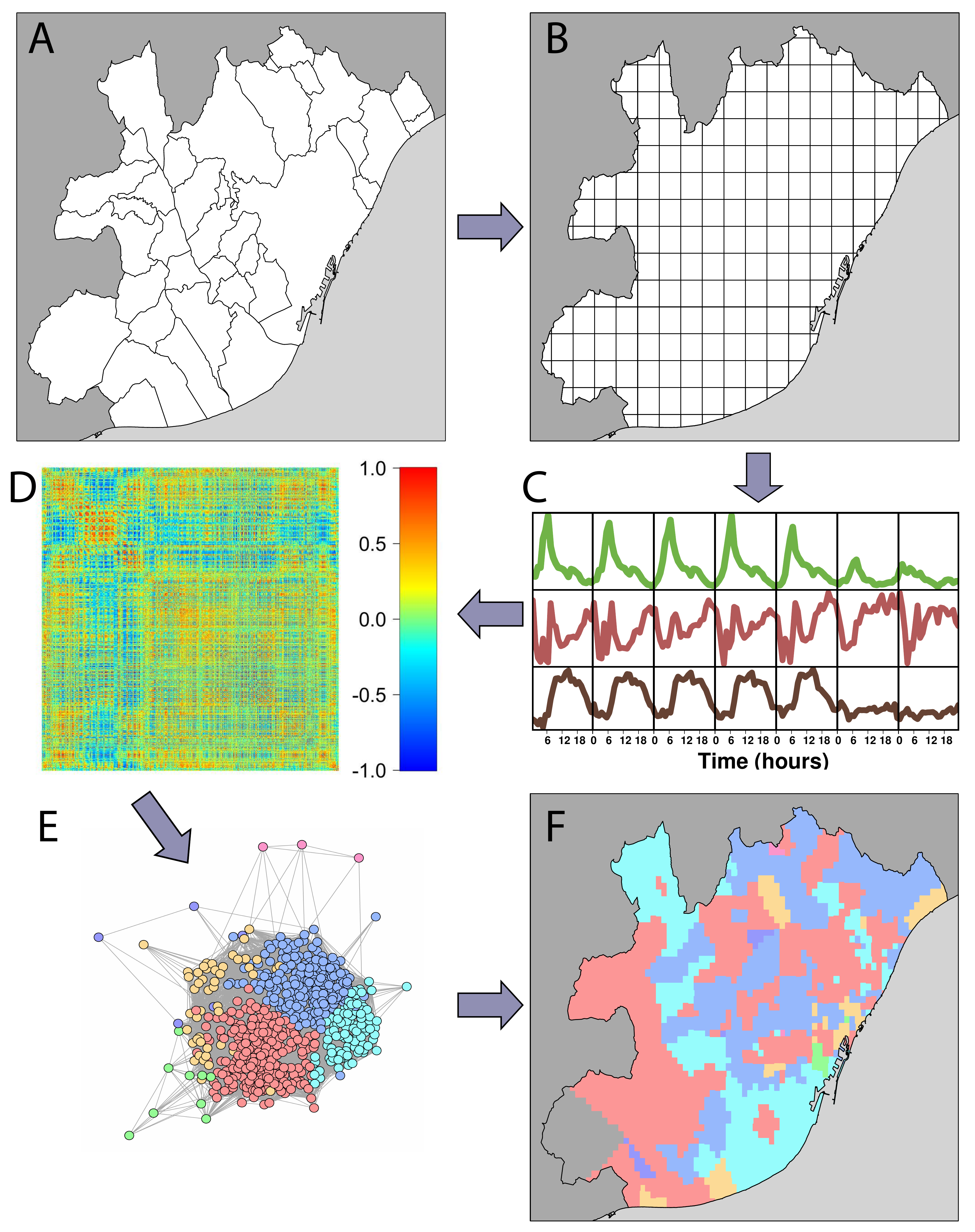}
\caption{\sf \textbf{Steps of the method to detect land use.} (A-B) The urban area is divided in cells of equal area. (C) For each cell, we calculate an activity profile in terms of phone calls along time during the days of the week. (D) A Pearson correlation matrix between cell activities is computed. Then the matrix formed by correlations over a threshold value $\delta$ is used to define an undirected weighted network (E), which is clusterized using community detection techniques and the results plotted again on the city map (F). \label{Fig1}}
\end{center}
\end{figure}

\vspace*{-0.5cm}
\section*{MATERIALS AND METHODS}

\subsection*{A network approach to detect land use}

Our database is composed of aggregated and anonymized call records during $55$ days between September and November $2009$ in Spain. Every time a user receives or makes a call, the event is registered together with the tower (BTS) providing the service. The positions of the BTSs are geo-referenced and so the activity levels of each spatial area can be tracked in time. For this work, we select the five most populated metropolitan areas of Spain: Madrid (with a population over $5.5$ million people), Barcelona ($3.2$ million), Valencia ($1.5$ million), Seville ($980,000$) and Bilbao ($900,000$). There is no unique definition for the border of an urban area. It may refer, for instance, to official, census or economic delimitation of the cities. Since the focus here is on urban land use, we are interested in identifying the inner zones of each city and, therefore, we use the definition of the metropolitan transportation offices: only areas served by metro or urban buses are considered. This is, nevertheless, an important question because the selection of borders may influence the scaling analysis when comparing across cities \cite{Arcaute2015,Louf2014b}. 

The space of the urban areas is divided following a Voronoi tessellation with the BTS location as centers. The extension of the areas served by each BTS is very different, since it depends on the expected peaks of demand. To ensure a common geographical framework, the five urban areas are divided in a grid with square cells of $500 \times 500\,m^2$ to which the activity is mapped. This should prevent spurious effects due to the Voronoi areas heterogeneity (see the Appendix for a detailed description of the cities and the division process). 

The activity (number of users) in each cell is monitored in time and then processed as illustrated in Figure \ref{Fig1}. Average activity profiles are estimated over each day of the week hour by hour in every cell. These profiles are normalized by the total hourly activity to subtract the trends introduced by the circadian rhythms. A Pearson correlation coefficient is then calculated between the activities of every pair of cells, obtaining a correlation matrix describing the level of similarity between activity profiles. The correlations can take positive and negative values. Distributions of these values are shown in the Figure S\ref{S4} of the Appendix. In order to remove non-significant and negative correlations we only consider Pearson correlation coefficients higher than a threshold $\delta$. As a result, we obtain one weighted network per urban area. We first note that variations of the threshold do not produce significant changes in the properties of the resulting network. The results in the main text refer to a value of $\delta$ equal to the correlation distribution dispersion. 

\begin{figure}[!h]
\centerline{\includegraphics[width=8.6cm]{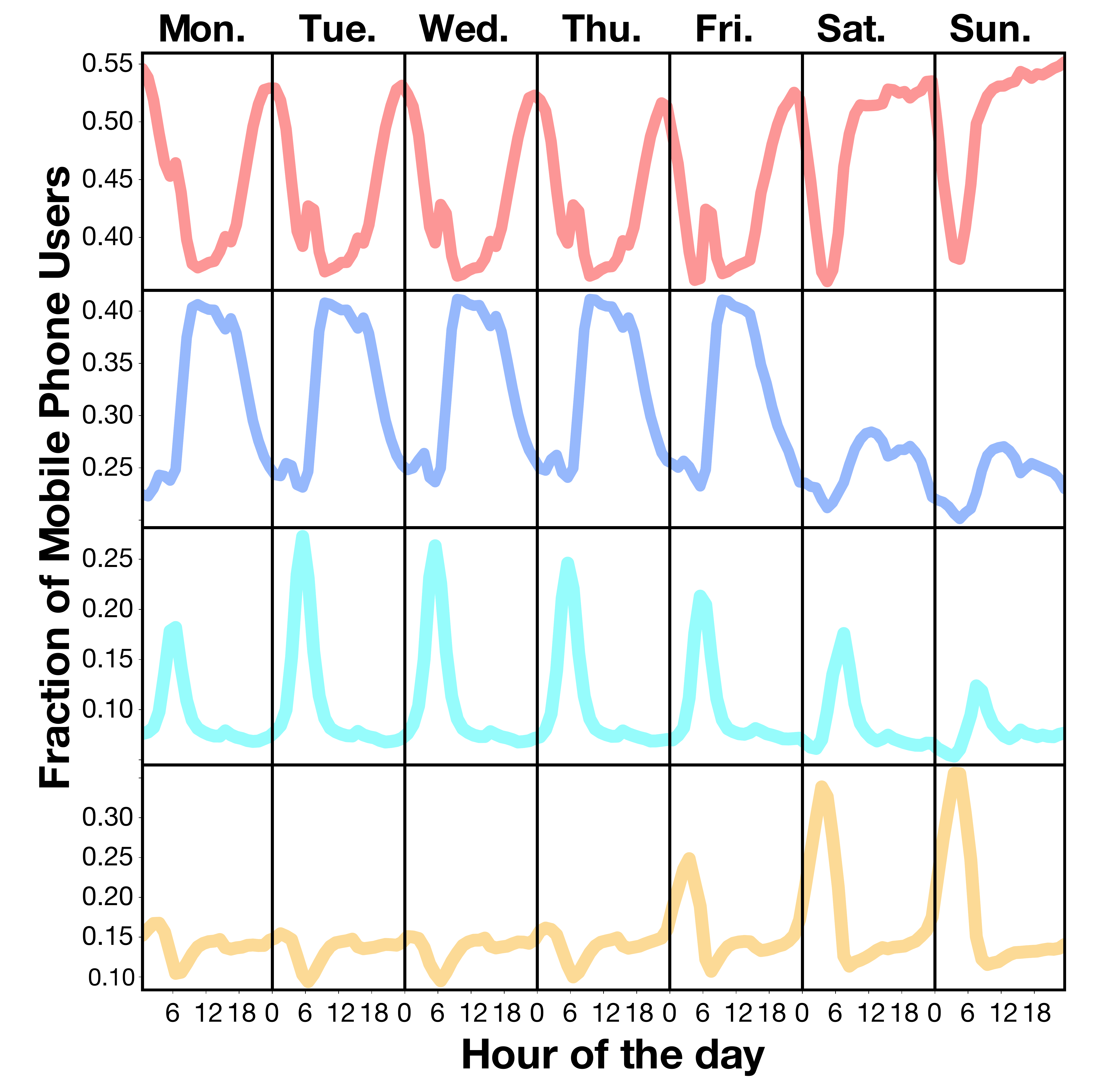}}
\caption{\sf \textbf{Temporal patterns associated with the four clusters for the metropolitan area of Madrid.} In red: Residential cluster; In blue: Business; In cyan: Logistics/Industry; And in orange: Nightlife.\label{Fig2}}
\end{figure}

Once the networks are built, their mesoscopic structure is analyzed using clustering techniques. The main advantage of community detection algorithms in networks compared to more classical clustering techniques based on dissimilarity matrix is that the number of clusters do not need to be fixed \textit{a priori}. However, it is important to note that different clustering methods can lead to distinct partitions of the networks. We report next results obtained with Infomap \cite{Rosvall2008}, while a systematic comparison with results obtained with other clustering tools is provided in the Appendix. As mentioned previously, Infomap does not require the input of a predetermined number of clusters. Therefore, it is interesting to find that in the five cities, between $98$ and $100\%$ of the cells are covered with only $4$ groups. Figure \ref{Fig2} shows how the activity looks like for each of these four clusters in Madrid (similar plots for Barcelona, Valencia, Seville and Bilbao are included as Figure S\ref{S9} and S\ref{S10}). 

Each of the clusters can be associated with a main land use: 

\begin{enumerate}

\item \textbf{Residential (red)}, which is characterized by low activities from $8$am to $5-6$pm. For the cells composing this group, the activity peaks around $7-8$am and during the evening. In the weekend, the activity is almost constant except for the night hours; 

\item \textbf{Business (blue)}, where the activity is significantly higher during the weekdays than during the weekends. Furthermore, it concentrates from $9$am to $6-7$pm. This land use designation can be related to a wide range of commercial, retail, service and office uses; 

\item \textbf{Logistics/Industry (cyan)}, where, as for Business, the activity is higher during the weekdays. We observe a large peak between $5$am and $7$am followed by a smaller peak around $3$pm. This cluster can be related to transport and distribution of goods: for example, "Mercamadrid" (the largest distribution area of Madrid) belong to this cluster; 

\item \textbf{Nightlife (orange)}, which is characterized by high activity during the night hours ($1$am-$4$am), especially during the weekends. During the weekdays, these areas show higher activity between $9$am and $6$pm, as for the Business cluster, which may be hinting a certain level of mixing in the land use. Some examples of this category are the "Gran Via" in Madrid and the "Ramblas" of Barcelona where abound theatres, restaurants and pubs mixed with offices and shops. This is typically the smallest cluster of the four in number of cells. 

\end{enumerate}

More systematically, we compared the land use patterns obtained with our algorithm with cadastral data. The dataset contains information about land use for each cadastral parcel of the metropolitan area of Madrid and Barcelona (about $650,000$ parcels). In particular, we have for each cadastral parcel the net internal area devoted to Residential, Business and Industrial uses. This information can be used to identify the dominant cadastral land use in each grid cell classified as Residential, Business and Industrial uses by the community detection algorithm. To do so we need to define a rule to determine what is the dominant land use in a cell. Intuitively, one would tend to identify the dominant land use in a cell as the land use class with the largest area. However, the nature of both land use assignations is very different: the cadastral data is based on the net internal area officially devoted to each activity and not necessarily on the number of people performing it. Therefore, Residential use is the land use class with the largest area in most of the cell leading to an over-representation of Residential cells in the metropolitan area. To circumvent this limitation we introduce two thresholds to identify Business and Logistics cells with cadastral data in order to obtain a distribution of the fraction of cells according to the land use type similar to the one obtained with our algorithm (see the Appendix for details). The overall agreement is high, we find a percentage of correct predictions equal to 65\% for Madrid and 60\% for Barcelona which is consistent with values obtained in other studies, $54\%$ in \cite{Toole2012} and $58\%$ in \cite{Pei2013}. Furthermore, for both case studies, almost all land use types have a percentage of correct predictions higher than $50\%$ (see the Appendix for details).

\section*{RESULTS}

\subsection*{Comparison of cities}

\begin{figure}[!h]
\centerline{\includegraphics[width=\linewidth]{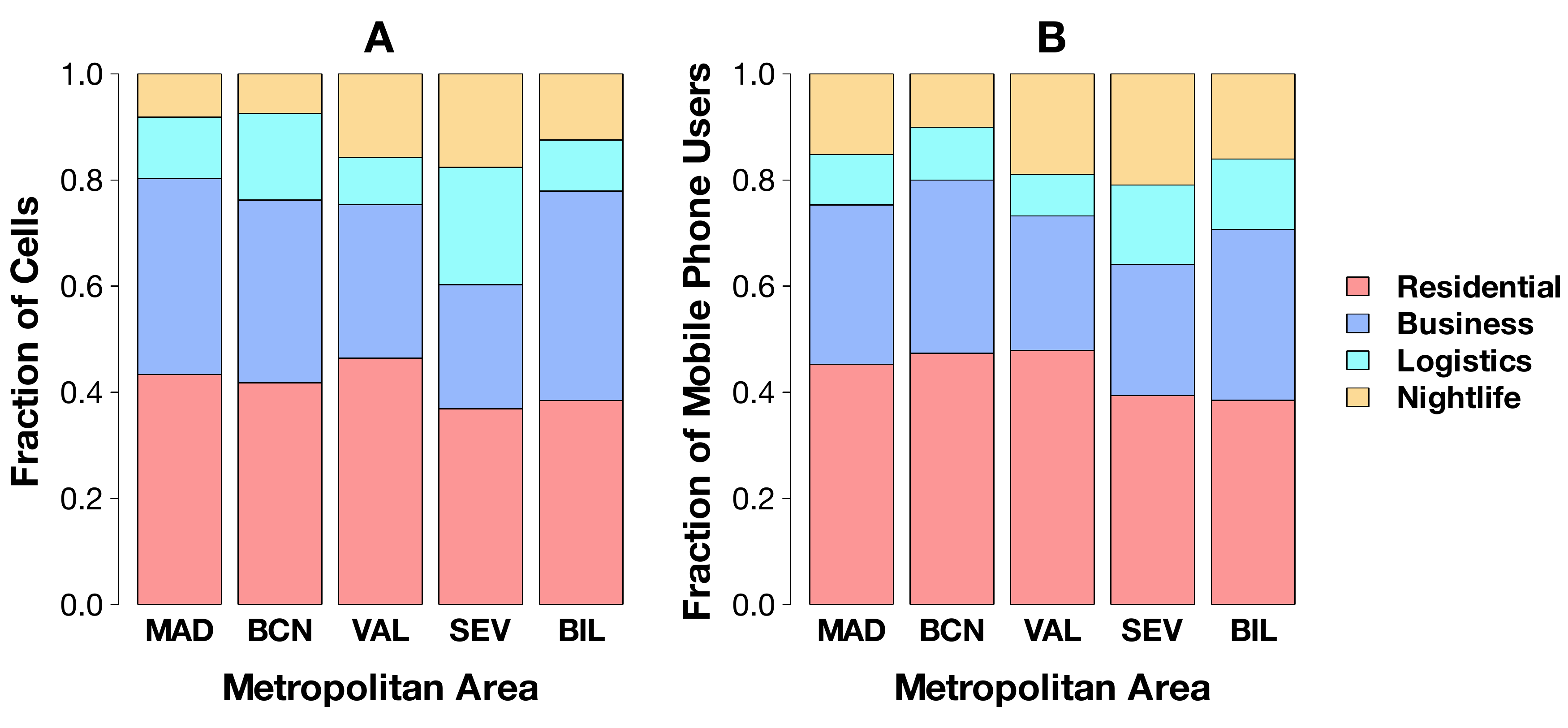}}
\caption{\sf \textbf{Fraction of cells (A) and mobile phone users (B) according to the type of land use for each case study.} The fraction of mobile phone users is averaged over the $168$ values of the time period. \label{Fig3}}
\end{figure}

Once defined the clusters, we can compare the proportion of land use type over the five case studies. Figure \ref{Fig3} shows the fraction of cells and the fraction of mobile phone activity averaged over the time period according to the land use patterns for each metropolitan area. We find similar results for the five case studies, the Residential land use type represents in average about 40\% of the cells and the mobile phone activity against 30\% for the Business category and less than 15\% for the Logistics and Nightlife clusters.

\begin{figure*}
\centerline{\includegraphics[width=11cm]{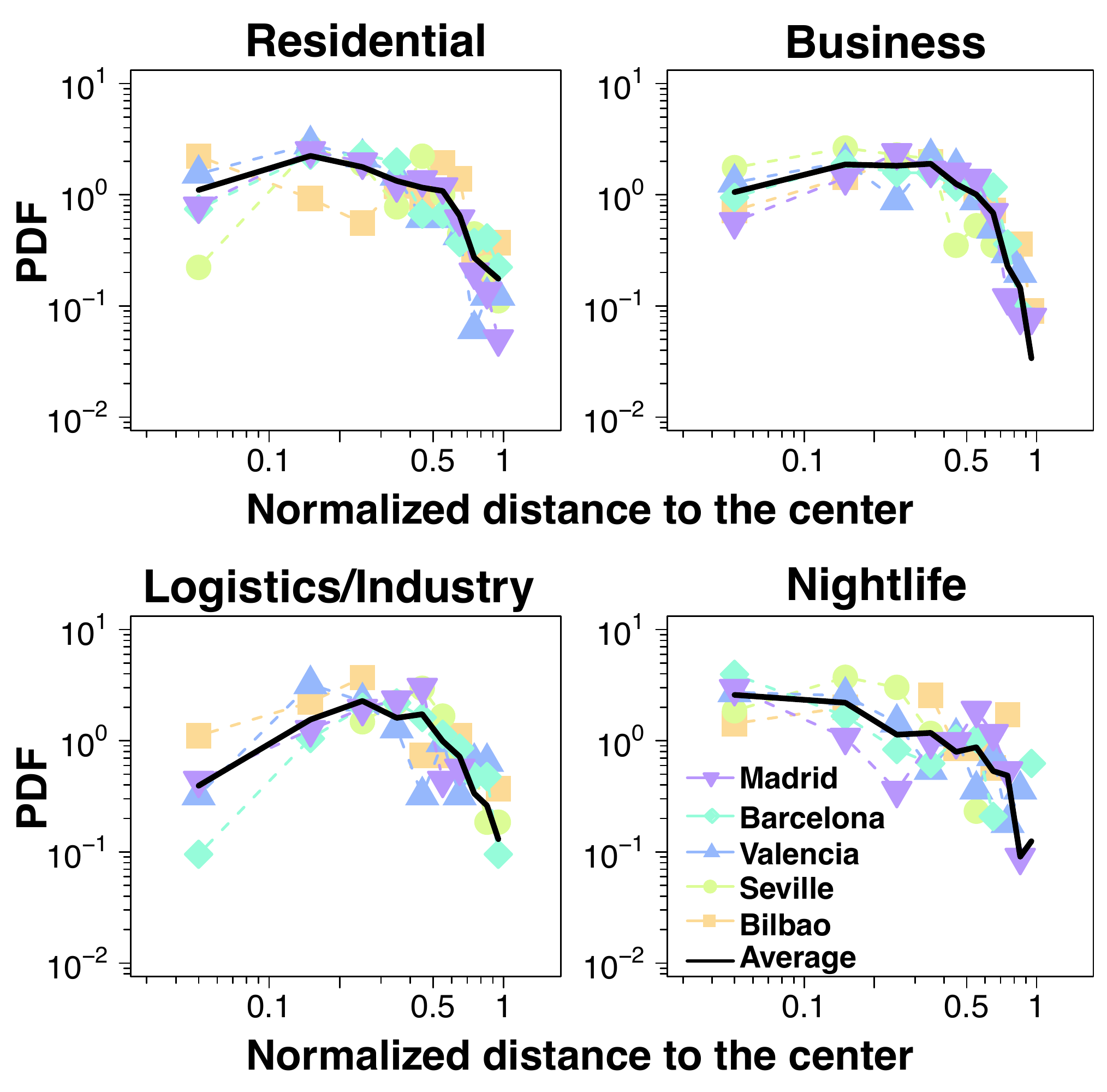}}
\caption{\sf \textbf{Distribution of the distance between the cells and the City Hall according to the type of land use.} The distance has been normalized by the maximum distance in each city.\label{Fig4}}
\end{figure*}

We can also study how the cells in each cluster are organized in the city's space. For the sake of comparison, we arbitrarily consider as city center the location of the City Hall and build a histogram with the number of cells at a certain distance from it. Since each city has a different spatial extension, distances are normalized by dividing by the maximal distance in each city so as to produce comparable results. The distributions are shown in Figure \ref{Fig4}, where average curves over all cities have been superimposed. It is interesting to note certain similarity in the distribution of cells for all urban areas. City size acts as a natural cutoff in the distributions, although no simple functional shape is found in any of the clusters. Residential cells are well distributed across the cities but with a maximum not very far from the center. Business cells appear at a similar distance as Residential but peaking a little further. Logistics and Industry are preferentially located in the periphery, while the Nightlife cells are well distributed along the urban areas but slightly more concentrated in the central areas. 

\begin{figure*}
\begin{center}
\centerline{\includegraphics[width=\linewidth]{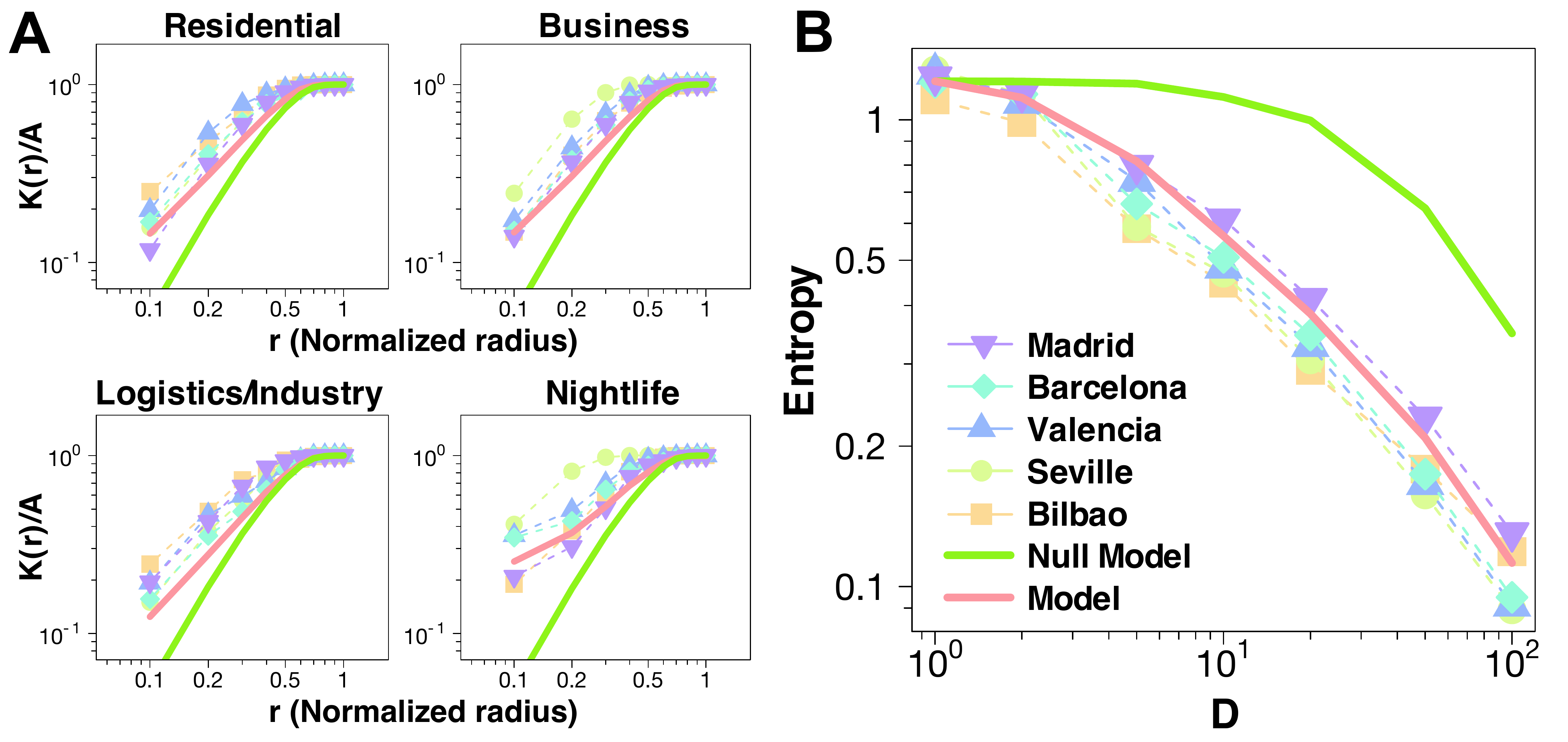}}
\caption{\sf \textbf{Comparison of the observed and the simulated Ripley's K and average entropy index.} (A) Ripley's K divided by the city area as a function of the search radius. The radius has been normalized with the maximum value in each urban area. (B) Average entropy index as function of the lateral number of divisions (inverse scale) D. The color and symbols of the curves represent different cities. The red curve corresponds to our model results and the green curve is the outcome of a random null model. Results for our model were obtained with a calibrated value of $\gamma=0.8$. The red and green curves display the average over $100$ realizations.\label{Fig5}}
\end{center}
\end{figure*}

In order to quantify land use distribution patterns, we use the Ripley's K \cite{Ripley1976} defined as

\begin{equation}
K(r) = \frac{A}{n^2} \sum_i^n N_i(r),
\end{equation}

where $A$ is the city area, $r$ the search radius (a geographical scale), the index $i$ runs over the cells in the urban area and $n$ is the total number of cells. $N_i(r)$ stands for the number of cells of a given type within a distance $r$ from the cell $i$. This indicator measures the spatial heterogeneity of a given type of cells. The baseline for homogeneous random systems is a growth $K(r)=\pi r^2$ until reaching $A$. If the value of $K(r)$ is over the random curve for a certain $r$ it implies that the system is clusterized at that scale. Since cities have different sizes, both $K(r)$ and the radius must be normalized by their maximum values ($A$ for $K(r)$ and the maximum distance for $r$). Curves for the normalized Ripley's K for each city and land use type are displayed in Figure \ref{Fig5}A as a function of the normalized radius. The $K(r)/A$ for each city are always above the green curve corresponding to a random distribution of land use types, indicating coarsening of land use. We find a scaling-like curve for all the land use types with most of the cities following well the general trend with some small deviations for Nightlife in Seville. 

Deepening the analysis, we can also define an entropy index to characterize the land use spatial organization. Let us consider a frame containing the full urban area, which is, in turn, sub-divided in a certain number $D^2$ of equal divisions. Each of these subdivisions, $B_i$, intersects the elementary cells so a certain fraction of area falls in each of the land use types: $f_i^R$ in the Residential cluster, $f_i^B$ in Business, $f_i^L$ in Logistics, and $f_i^N$ in Nightlife. An entropy index, $E_i$, can be defined for $B_i$ as

\begin{equation}
E_i=-\sum_\alpha f_i^{\alpha} ln(f_i^{\alpha}),
\end{equation}

where $\alpha$ runs over the four clusters. The entropy $E_i$ is then averaged over all the divisions to obtain a global metric for the city at a given scale $E(D)$. $E(D)$ tends to zero if the land use within the divisions becomes unique, as occurs for instance at large $D$ (small spatial scales). On the other extreme, when $D\rightarrow 1$ , $E(D)$ converges to a fixed value describing the full city. Figure \ref{Fig5}B shows how the average entropy behaves with $D$. The curves are similar across cities, recalling the shape of scaling functions. This is not surprising if the concept of a fractal-like distribution of the city activity applies as has been previously discussed in the literature \cite{Bettencourt2007,Batty2008,Bettencourt2013,Bettencourt2010,Batty2013,Arcaute2015}.

\begin{figure*}
\begin{center}
\centerline{\sf \includegraphics[width=11cm]{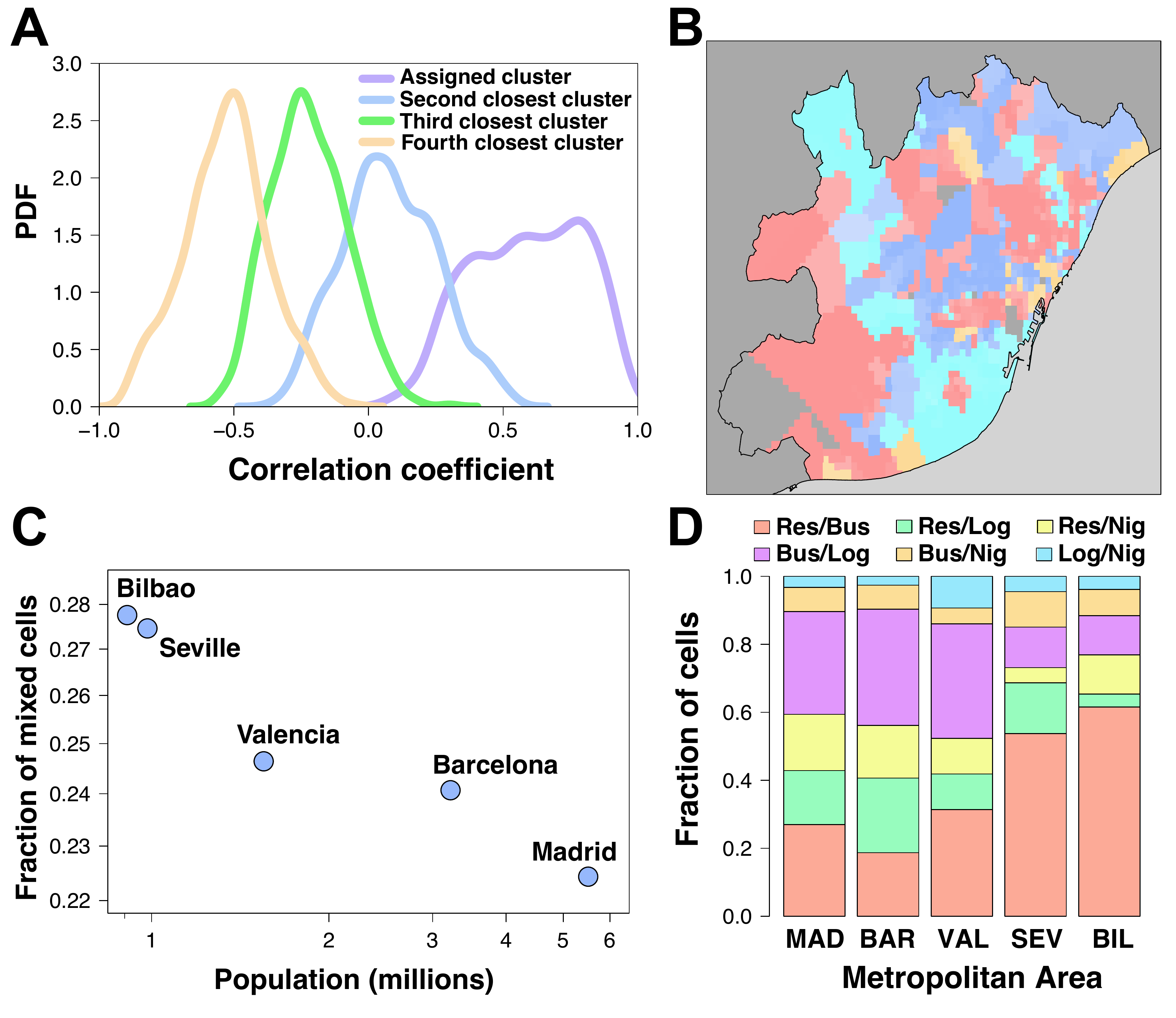}}
\caption{\sf \textbf{Land use mixing.} (A) Distribution of the Pearson correlations between cells activity and the average cluster profiles. (B) Map of Barcelona displaying the four clusters with the colors varying from white to the baseline according to the intensity of the relation with the assigned cluster of each cell. The color code is red for Residential, blue for Business, cyan for Logistics/Industry and orange for Nightlife. (C) Fraction of mixed cells as a function of the city population. (D) Fraction of cells classified by the type of land use mixing among those with two types of land use.\label{Fig6}}
\end{center}
\end{figure*}

\subsection*{Modeling land use}

Urban land use models in the literature are typically built with relative elaborated mechanisms \cite{Wilson1969,Decraene2013}. If basic in the rules, the models typically refer to characteristics of cities such as the population or activity distributions \cite{Makse1998,Louf2013,Louf2014}. The shape of $E(D)$ can be explained, however, by a simple model inspired by Schelling's segregation \cite{Schelling1971}. It is important to stress that this model is not intended to reproduce all the processes leading to the land use formation, but to explain the scaling of its spatial distribution patterns.    

The basic framework is a lattice in $2$D representing the urban space. Initially, a variable $t_i$ with a land use type is assigned to every cell $i$ at random (Residential $R$, Business $B$, Logistics $L$ or Nightlife $N$). The global fraction of cells of each type respects the proportions found in the empirical data in such a way that $E(D=1)$ coincides with the observations. A {\em satisfaction} index, $S_i$, is then defined per cell taking into account its type and those of its neighbors. Similarly to Schelling's model, we assume that the satisfaction increases when a cell is surrounded by cells of its own type. Otherwise, $S_i$ depends on the particular combinations of types. Some land uses \textit{attract} each other as, for instance, Residential and Business, while others repel as Residential and Logistics. The existence of rules of attraction and repulsion between different types of land use has already been considered in the past like for example in \cite{Krugman1996,Feng2011}. To be specific if $p_t^i$ is the fraction of neighbors of $i$ of type $t$, then  $S_i$ is calculated as

\begin{equation}
\begin{array}{ll}
\mbox{if }t_i = L,  & S_i = \delta_{p_L^i,1} , \nonumber\\ $\,$ \\
\mbox{if }t_i = N, & S_i = p_N^i \, \delta_{p_L^i,0} , \\
$\,$ & $\,$\\
\mbox{if }t_i = R,B, & S_i = \left\{ \begin{array}{l}  \delta_{p_L^i,0} \mbox{ with probability } \gamma, \\ $\,$ \\
                                                  p_{R,B}^i \, \delta_{p_L^i,0} \mbox{ with probability } 1-\gamma, 
                         \end{array} \nonumber \right.
\end{array}			 
\end{equation}

where $\delta_{p,x}$ is the Kronecker delta (equal to one if $p = x$, zero otherwise) and $\gamma$ is the only model parameter. Note that the first condition implies that for Logistic cells $S_i = 1$ only if they are surrounded by cells of the same type, and that cells of other types have zero satisfaction if surrounded by any Logistic one. With this rule, we introduce a tendency to locate Industry and Logistics out of the core areas of the cities. Residential and Business cells have a certain tolerance to the $R, B$ and $N$ types with $\gamma$ acting as a mixing control parameter: if $\gamma = 0$, mixing is not favored. A global satisfaction measure is defined as the sum over all the cell satisfaction indices, $S = \sum_i S_i$. The model is updated by choosing random pairs of cells and interchanging their land use if the exchange increases $S$. This process is repeated until the satisfaction reaches a stationary state. 

Calibrating the single parameter $\gamma$, we can reproduce the observed $K(r)/A$ and $E(D)$ scaling in the real urban areas (see red curves in Figure \ref{Fig5}). The value of the mixing parameter at which the best average results are obtained is $\gamma = 0.08$ (Figure S\ref{S8}).  For comparison sake, we have included a null model in which the land use types are distributed at random, keeping the real proportions (green curves in Figure \ref{Fig5}). Unsurprisingly, the null model fails at reproducing the curves obtained with the data, mainly because, generally, areas of a given land use type tend to cluster together to form land use zones. More interestingly, assuming that the satisfaction of a cell increases when it is surrounded by cells of its own type and that Logistic cells and cells repel each other (i.e. $\gamma = 0$) is also insufficient to reproduce the properties observed in the data (for more details see Figure S\ref{S8} in Appendix). Therefore, taking into account the attraction between Residential and Business areas seems to be crucial for the reproduction of land use spatial organization in cities.

\subsection*{Mixing of land use types}

So far, we have considered that each elementary cell has a unique land use type associated. This condition can be easily relaxed. If an average activity profile is defined for each of the four clusters, a Pearson correlation coefficient between the activity profile in each cell and the clusters' averages can be calculated. The distribution of correlation values is shown in Figure \ref{Fig6}A. The highest correlation value corresponds typically to the cluster at which the cell is assigned. Still, in some cases, positive correlation values are found for other or even two other clusters. For every cell, 
we can quantify the intensity of its relation with each cluster by summing over these positive correlations and normalizing by the total. A map of the Barcelona metropolitan area with the intensity of each cell relation with its assigned cluster is shown in Figure \ref{Fig6}B. The colors represent the four main type of cells and the color saturation is related to the correlation: darker if the correlation is high, paler otherwise. Most cells match well with their original assigned cluster, keeping darker colors, while some are brighter, implying a higher level of land use mixing. 

We arbitrarily define a cell as mixed when the normalized correlations fall within the interval $0.3-0.7$ for other clusters besides the assigned one. The fraction of mixed cells as a function of the city population is displayed in Figure \ref{Fig6}C. Larger cities show lower mixing and contain areas devoted to more specific purposes. Figure \ref{Fig6}D illustrates how the land use types combine in each cell. Business and Residential integrate well together as in our model, increasingly so for smaller cities. The mixing proportions are city-dependent and act as a fingerprint to characterize each urban area. This feature may be used to classify cities with similar land use mixing patterns. Besides population size, causal links between level of spatial mixing and city shape, area, age of the city, function, etc. will be explored in future investigations. The mixing proportion can be either related to the organization of cities as monocentric or polycentric \cite{Louail2014,Roth2011}. Smaller cities display a more monocentric structure, which can be associated with the mixing of land use types given the most restricted space. 

\section*{DISCUSSION}

In summary, we introduce a method to automatically detect land use from electronic records and apply it to the five largest urban areas of Spain in order to perform a systematic comparison across them on the land use distribution. The urban space is divided in a regular grid to prevent geographic heterogeneity and to maintain the spatial scale under control. The user activity profiles are monitored in each unit cell along time, and then a correlation matrix is established between the profiles of every pair of cells. This correlation matrix encodes the functional network of each city. We analyze them by using network clustering techniques, which ensures that cells showing similar use profiles are grouped together. This method has been applied to the five most populated Spanish cities: Madrid, Barcelona, Valencia, Seville and Bilbao. Since the delimitation of urban areas could affect the results, the definition of the municipal transport offices is employed in each case. Interestingly given that the method is unsupervised, four groups consistently appear as dominant in all cities. They correspond to activity profiles compatible with main land uses in Residence, Business, Logistics/Industry and Nightlife. Not only the types are the same across cities, but also the proportions of cells and area devoted to each type are similar. 

We also study the distribution of the four land use types at different spatial scales. We define the Ripley's K and the  entropy index for each land use type and the behavior of both metrics is explored as the spatial scale varies from the full city (macroscopic scale) to a single cell (microscopic). The five cities show similar scaling curves for the metrics, implying comparable structures regarding how the four types amalgamate at the urban level. The shape of the scaling curves can be explained by a simple model that has been proposed in this work. The model is based on a Schelling-like segregation in which the different land use types interact to generate a spatial distribution in the city. Cells in a given land use type tend to maximize the number of neighbors undergoing equivalent uses. This rule induces a tendency to coarsening in land use types. The different land uses interact by attracting each other, such as services and residential areas, or by repelling like industry and almost any other type. The calibration of a single parameter regulating the intensity of the attraction between services, residential uses and nightlife is enough to reproduce the scaling curves observed in the real cities. Moreover, we also demonstrate that a model without land use type interactions cannot recreate the empirical scaling. 

Similarities across cities break down when one focuses on how the land use types mix microscopically within each unit cell. A characteristic mixing profile is detected for every urban area, providing an individual city fingerprint. Further data on other cities could help to elucidate whether different typologies exist at this microscopic mixing level. In conclusion, despite   further data from other countries and sources could be important to confirm our results, we find that a coherent picture emerges in the land use organization of major urban areas and that its origin can be explained with a basic model.

\vspace*{0.5cm}
\section*{ACKNOWLEDGEMENTS}
Partial financial support has been received from the Spanish Ministry of Economy (MINECO) and FEDER (EU) under projects MODASS (FIS2011-24785) and INTENSE@COSYP (FIS2012-30634), and from the EU Commission through projects EUNOIA, LASAGNE and INSIGHT. The work of ML has been funded under the PD/004/2013 project, from the Conselleria de Educaci\'on, Cultura y Universidades of the Government of the Balearic Islands and from the European Social Fund through the Balearic Islands ESF operational program for 2013-2017. JJR from the Ram\'on y Cajal program of MINECO. 

\bibliographystyle{unsrt}
\bibliography{FN}

\makeatletter
\renewcommand{\fnum@figure}{\small\textbf{\figurename~\textbf{S}\thefigure}}
\renewcommand{\fnum@table}{\small\textbf{\tablename~\textbf{S}\thetable}}
\makeatother

\setcounter{figure}{0}
\setcounter{table}{0}
\setcounter{equation}{0}

\section*{APPENDIX}

\subsection*{Case studies}

\begin{figure}[b]
  \begin{center}
		\includegraphics[scale=0.5]{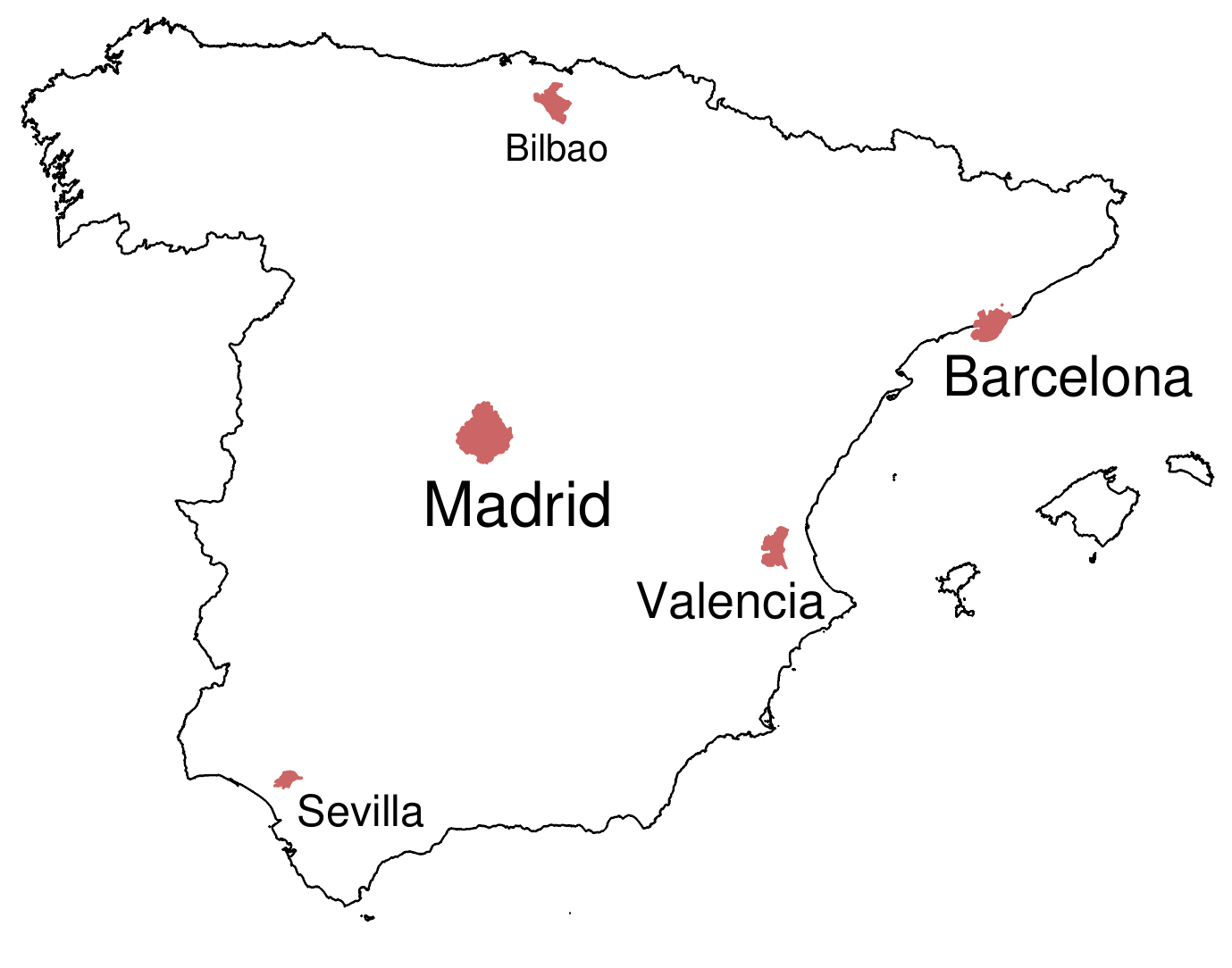}
	\end{center}
		\caption{\sf \textbf{Map of the metropolitan areas.} \label{S1}}
\end{figure}

In this study, we focused on the five biggest metropolitan areas of Spain, Madrid, Barcelona, Valencia, Seville and Bilbao (Figure S\ref{S1}). These metropolitan areas are very different in terms of sizes and populations (Table S\ref{MA}). For all cities we have selected as urban area the one served by public transportation (bus and metro) instead of the official definition that in the case of Seville includes a much larger extension relatively depopulated.  

\begin{table*}
	\caption{\sf \textbf{Summary statistics on the metropolitan areas.}}
	\label{MA}
		\begin{center}
			\begin{tabular}{lccc}
				\hline
				 \centering Metropolitan area & Number of municipalities & Number of inhabitants  & Area (km$^2$)\\
				\hline
				Madrid    & 27 & 5,512,495 & 1,935.97  \\ 
				Barcelona & 36 & 3,218,071 & 634      \\      
				Valencia  & 43 & 1,549,855 & 628.81  \\ 
				Sevilla   &  8 & 983,852   & 352  \\ 
				Bilbao    & 34 & 908,916   & 500.2  \\ 
				\hline
	  	\end{tabular}
	  \end{center}
\end{table*}

\subsection*{Data pre-processing}

Mobile phone records of anonimyzed users during 55 days (hereafter noted $T$) within the period of September-November 2009 were aggregated in two different ways. The aggregated data corresponds to the number of users per hour and per base transceiver stations (BTSs) identified with UTM (WSG84) coordinates. A user may appear connected to more than one BTS within a period of one hour. To avoid over counting people the following criteria was used when aggregating the data: each person shall count only once per hour. If a user is detected in $k$ different positions within a certain 1-hour time period, each registered position will count as ($1/k$) "units of activity". From this aggregated data activity per BTS and per hour is calculated for each day. In order to compute the number of mobile phone users $P_{g,d}(h)$ in a grid cell $g$ (dimension $0.5\times 0.5$ km$^2$) for a day $d\in T$ between $h$ and $h+1$, where $h \in |[0,23]|$, we first computed the Voronoi cells associated with each BTS.

\begin{figure*}
    \begin{center}
		   \includegraphics[width=\linewidth]{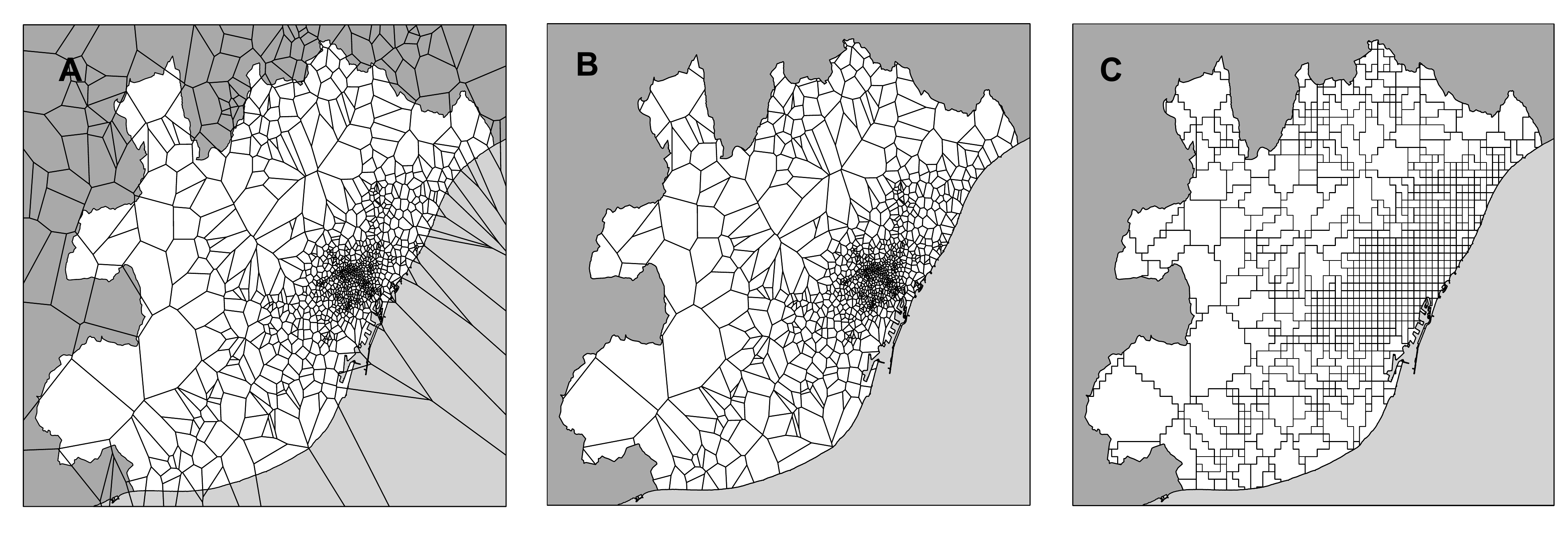}
		\end{center}
		\caption{\sf \textbf{Map of the metropolitan area of Barcelona.} The white area represents the metropolitan area, the dark gray area represents territory surrounding the metropolitan area and the light grey area represents the sea. (A) Voronoi cells of the mobile phone antennas point pattern. (B) Intersection between the Voronoi cells and the metropolitan area. (C) Recording sites composed of grid cells of dimension $0.5\times 0.5$ km$^2$. \label{S2}}
\end{figure*} 

\begin{figure*}
    \begin{center}
		   \includegraphics[width=\linewidth]{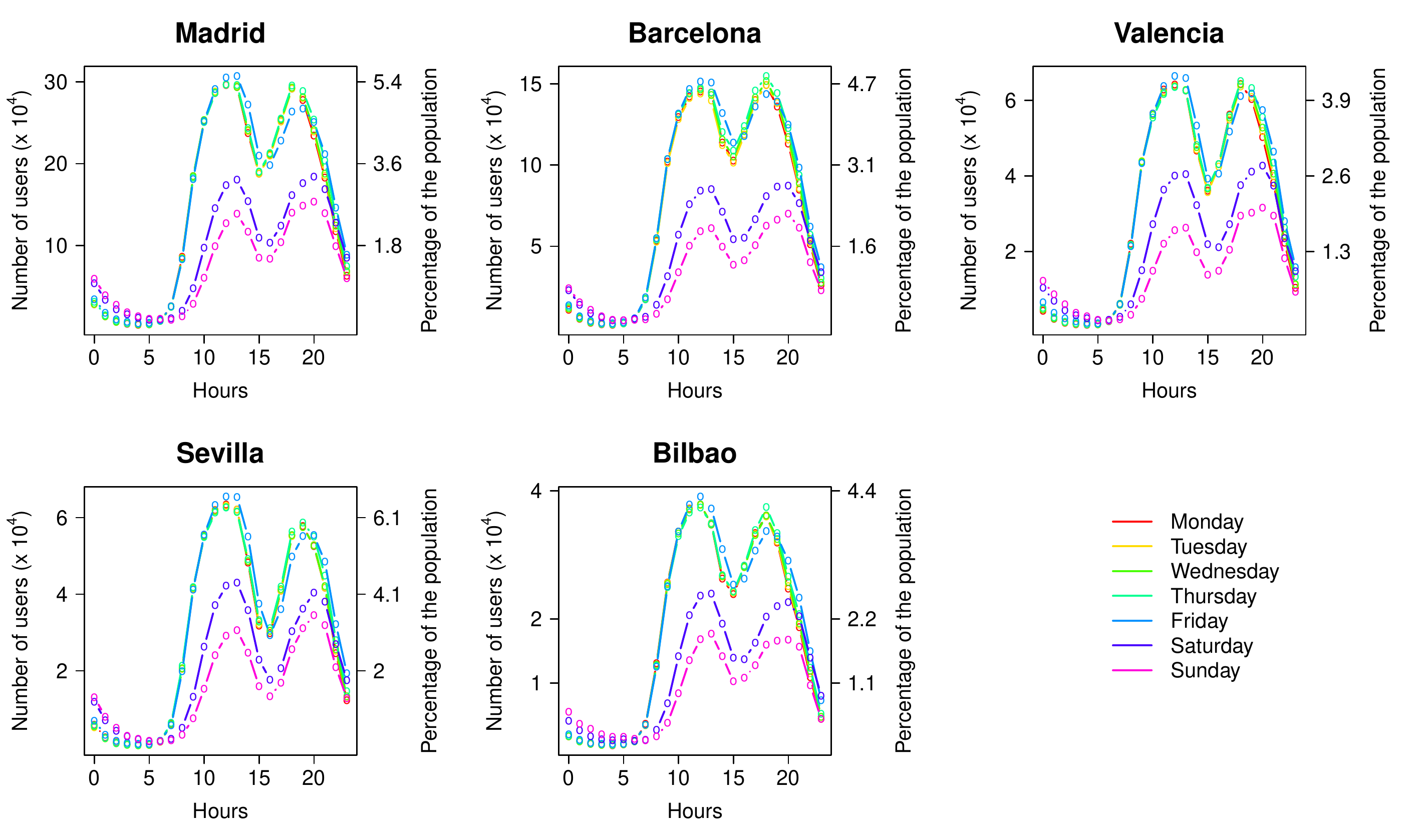}
		\end{center}	
		\caption{\sf \textbf{Average number of mobile phone users per hour according to the day of the week for the five metropolitan areas.}\label{S3}}
\end{figure*}

\subsubsection*{Voronoi cells}

First we remove the BTSs with zero mobile phone users and we compute the Voronoi cells associated with each BTSs of the metropolitan area (hereafter called MA). We remark in Figure S\ref{S2}A that there are four types of Voronoi cells:

\begin{enumerate}
   \item The Voronoi cells contained in MA.
	 \item The Voronoi cells between MA and the territory outside the metropolitan area.
	 \item The Voronoi cells between MA and the sea (noted S).
	 \item The Voronoi cells between MA, the territory outside the metropolitan area and the sea.
\end{enumerate}

To compute the number of users associated with the intersections between the Voronoi cells and MA we have to take into account these different types of Voronoi cells. Let $m$ be the number of Voronoi cells (ie BTSs), $N_{v,d}(h)$ the number of users in a Voronoi cell $v$ (on day $d$ at time $h$) and $A_v$ the area of $v$, $v \in |[1,m]|$. The number of users $N_{v\cap MA,d}(h)$ in the intersection between $v$ and MA is given by the following equation:

\begin{equation}
	N_{v\cap MA,d}(h) = N_{v,d}(h) \left(\frac{\displaystyle A_{v\cap MA}}{A_v - A_{v\cap S}}\right)
	\label{vMA}
\end{equation}

We note in Equation S\ref{vMA} that we have removed the intersection of the Voronoi area with the sea, indeed, we assume that the number of users calling from the sea are negligible. Now we consider the number of mobile phone users $N_{v,d}(h)$ and the associated area $A_v$ of the Voronoi cells intersecting MA (Figure S\ref{S2}B).

\subsubsection*{Grid cells}

 Let $n$ be the number of grid cells, the number of mobile phone users $N_{g,d}(h)$ (on day $d$ at time $h$) is given by the following equation, $\forall$ g $\in |[1,n]|$:

\begin{equation}
	N_{g,d}(h)=\sum_{v=1}^m N_{v,d}(h) \frac{\displaystyle A_{v\cap g}}{\displaystyle A_{v}}.
	\label{Pgdh}
\end{equation}

Then the set of days $T$ is divided into subsets $T_{w} \subset T$ and the average number of mobile phone users is computed for each day of the week $w$ (Equation S\ref{Pgwh}). 

\begin{equation}
	N_{g,w}(h)=\frac{\sum_{d\in T_{w}}N_{g,d}(h)}{|T_{w}|}
	\label{Pgwh}
\end{equation}

The average number of mobile phone users for the metropolitan areas according to the time and the day of the week are plotted on Figure S\ref{S3}. The profile curve shows two peaks, one peak around 12AM and an other one around 7PM. It also shows that the number of mobile phone users is higher during weekdays than during weekend. 

$N$ is normalized such that the total number of users at a given time on a given day is equal to $1$, Equation S\ref{chapp},

\begin{equation}
	\hat{N}_{g_0,w}(h)=\frac{\displaystyle N_{g_0,w}(h) }{\sum_{g=1}^n \displaystyle N_{g,w}(h)}
	\label{chapp}
\end{equation}

This normalization allows for a direct comparison between sources with different absolute user's activity. For a given grid cell $g=g_0$ we defined the temporal distribution of users $\hat{N}_{g_0}$ as the concatenation of the temporal distribution of users associated with each day of the week. For each grid cell we obtained a temporal distribution of users (also called signal) represented by a vector of length $24 \times 7$. It is possible that some grid cells have exactly the same signal because some Voronoi cells may contain several cells, in this case the grid cells have been aggregated (Figure S\ref{S2}C). 

\begin{figure*}
  \begin{center}
		\includegraphics[scale=0.8]{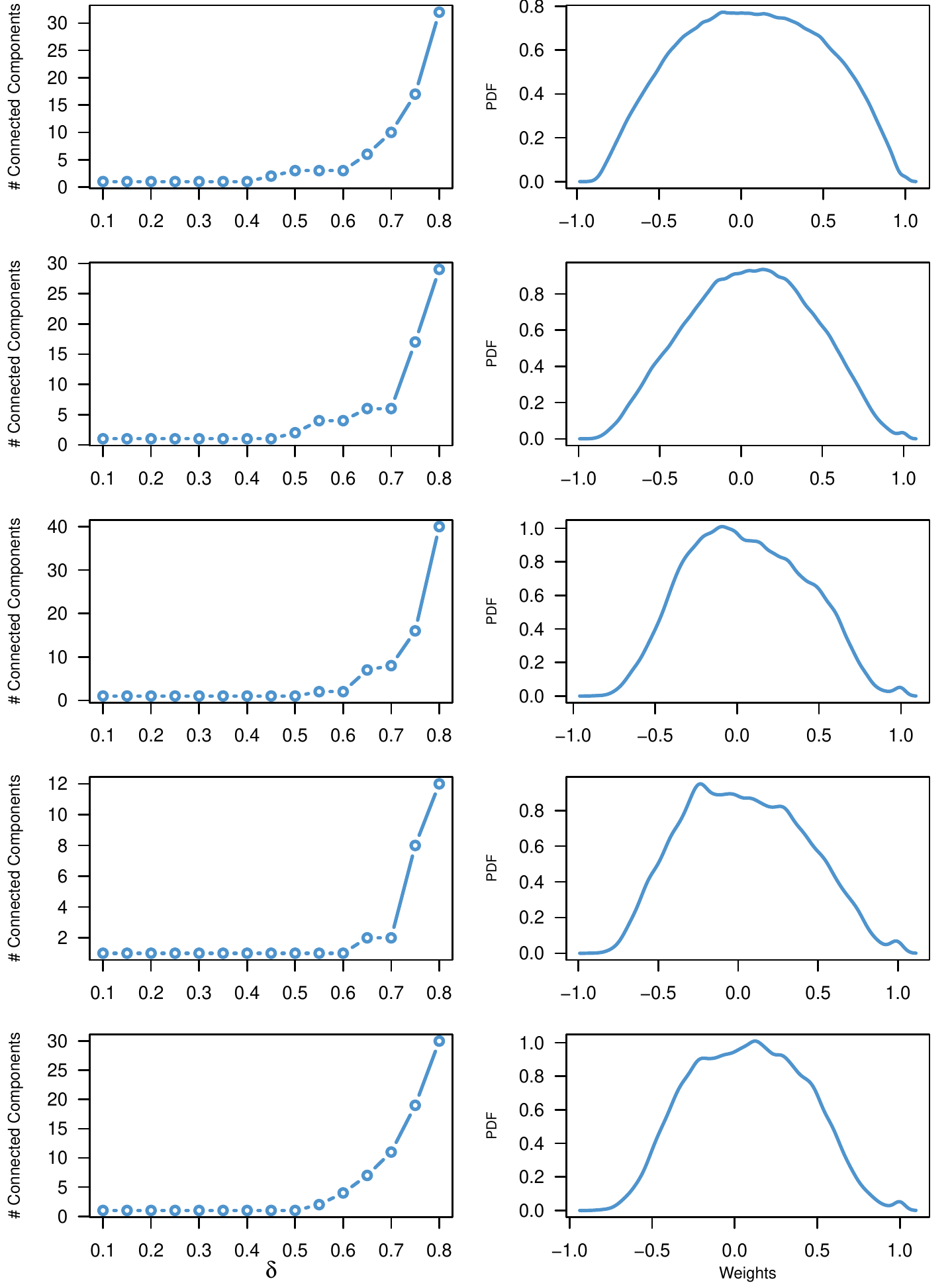}
	\end{center}
		\caption{\sf \textbf{Number of connected components as a function of $\delta$ (Left) and weight distribution (Right) for the five case studies.} From top to bottom, Madrid, Barcelona, Valencia, Sevilla and Bilbao. \label{S4}}
\end{figure*}

\subsection*{Functional network}

\subsubsection*{Choice of $\delta$}

In the method used to extract the functional network from the mobile phone data presented in the main text we apply a threshold $\delta$ to the correlation matrix in order to remove the noise and negative correlations from the correlation matrix. Hence, we have to choose a value of $\delta$ high enough to remove the noise but not too high in order to preserve the structure and the properties of the network. Figure S\ref{S4} displays the distribution of the weights (i.e. correlation coefficient) for the five case studies. One can observe that these distributions can be approximated by a Gaussian distribution. Therefore, we have decided to keep only edges with a weight higher than the weight distribution's standard deviation. In Figure S\ref{S4} we note that for $\delta$ lower than the weight distribution's standard deviation (around $0.4$, see details in Table S\ref{t}) the number of connected components is equal to 1.

\begin{table}[t]
  			  \fontsize{6}{6}\selectfont
	\caption{\sf \textbf{Statistical properties of the functional networks.}}
	\label{t}
		\begin{center}
			\begin{tabular}{lccccccccc}
				\hline
				 City & $SD$ & $N$ & $E$ & $<k>$ & $<k>/N$ & $C$  & $L$ & $C_r$  & $L_r$\\
				\hline
Madrid	&	0.42	&	1,381	&	222,227	&	321.8	&	0.233	&	0.69	&	2.04	&	0.31	&	1.77	\\
Barcelona	&	0.38	&	652	&	46,573	&	142.9	&	0.219	&	0.62	&	2.02	&	0.29	&	1.79	\\
Valencia	&	0.35	&	351	&	13,847	&	78.9	&	0.225	&	0.66	&	2.06	&	0.31	&	1.84	\\
Sevilla	&	0.38	&	188	&	3,700	&	39.2	&	0.209	&	0.62	&	2.15	&	0.26	&	1.81	\\
Bilbao	&	0.35	&	267	&	8,915	&	66.8	&	0.25	&	0.67	&	2.03	&	0.39	&	1.76	\\					
				\hline
	  	\end{tabular}
	  \end{center}
\end{table}

Table S\ref{t} summarizes the statistical properties of the functional networks obtained for the five case studies. In these tables we can observe the threshold ($SD$), the number of nodes (i.e number of cells) ($N$), the number of edges ($E$),  the average degree ($<k>$), the average clustering coefficient ($C$) and the average shortest path length ($L$). The average clustering coefficient $C_r$ and the average shortest path length $L_r$ have been obtained with a randomly rewired network preserving the degree of the original network by permuting links (4 x (number of edges) times) \cite{Maslov2002}. We observe that the five networks are very similar, characterized by a high clustering coefficient and low average shortest path.

\begin{figure*}
  \begin{center}
		\includegraphics[scale=0.5]{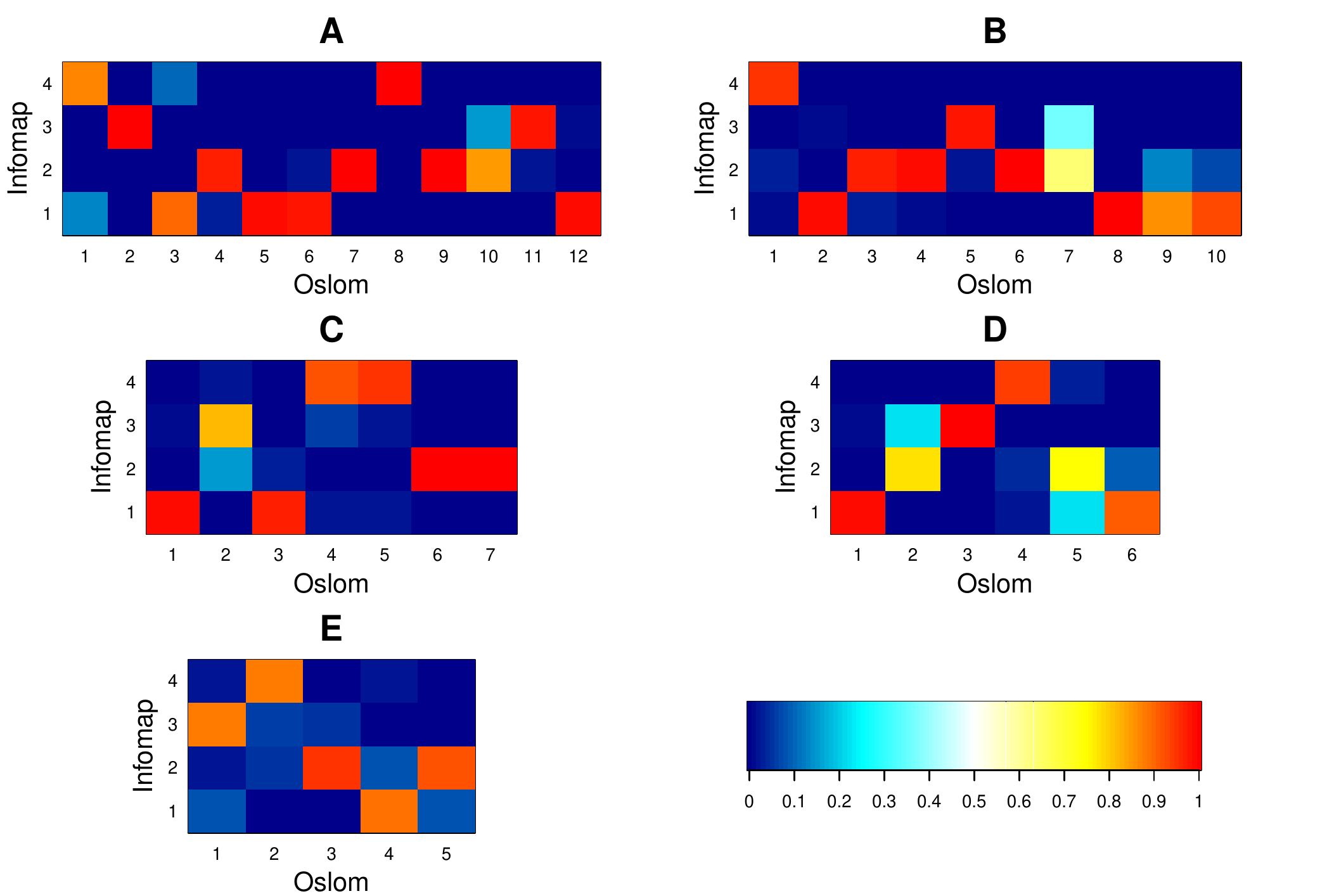}
	\end{center}
		\caption{\sf \textbf{Contingency tables between the partitions obtained with Infomap and OSLOM for each case study.} (A) Madrid. (B) Barcelona. (C) Valencia. (D) Sevilla. (E) Bilbao. Each row represents a cluster obtained with Infomap and each column represents a cluster obtained with Oslom. The matrices have been normalized so that the sum of each column is equal to one. \label{S5}}
\end{figure*}

\subsubsection*{Community detection}

Community detection in complex networks has recently been the subject of an abundant literature and a large number of algorithms has been proposed the last few years. The purpose of these algorithms is to identify closely connected groups of nodes within a network. To do so, several techniques are used such as maximizing the modularity, measuring probability flows of random walks or optimizing the local statistical significance of communities. 

In this paper, we have decided to use the Infomap method proposed in \cite{Rosvall2008}. Infomap finds communities by using the probability of flow of random walks on the network as a proxy for information flow in the real system and then decompose the graph into groups of nodes among which information flows easily. As shown in \cite{Lancichinetti2009}, this method gives good results, however, to evaluate the robustness of the results, the analysis has also been performed with two other clustering methods, Oslom \cite{Lancichinetti2010,Lancichinetti2011} and Louvain \cite{Blondel2008}. Oslom is a method based on a topological approach to detect statistically significant cluster whereas Louvain is based on modularity optimization which means finding the optimal partition maximizing the density of links within clusters and minimizing the density of links between clusters.

\begin{figure*}
  \begin{center}
		\includegraphics[scale=0.7]{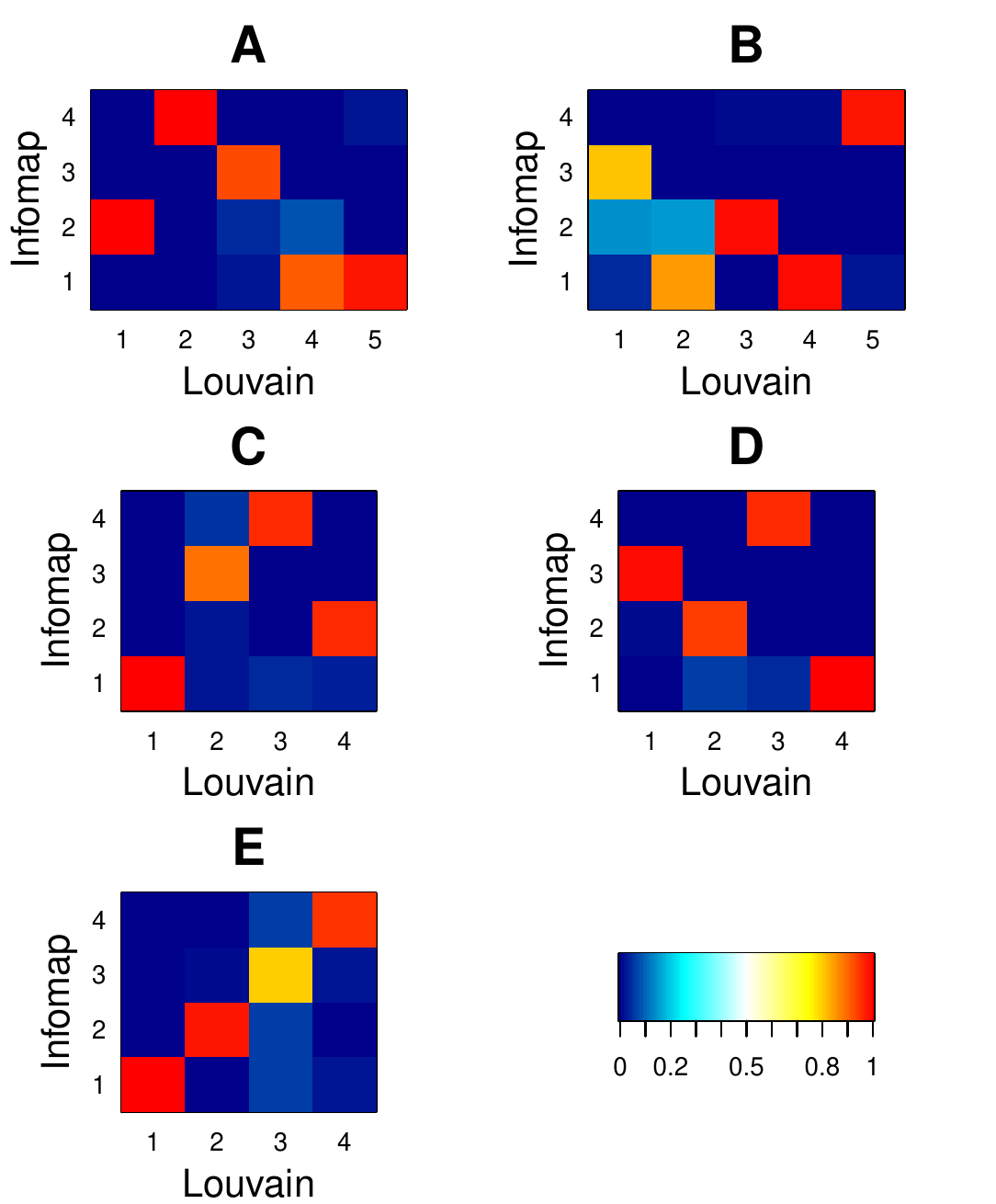}
	\end{center}
		\caption{\sf \textbf{Contingency tables between the partitions obtained with Infomap and Louvain for each case study.} (A) Madrid. (B) Barcelona. (C) Valencia. (D) Sevilla. (E) Bilbao. Each row represents a cluster obtained with Infomap and each column represents a cluster obtained with Louvain. The matrices have been normalized so that the sum of each column is equal to one. \label{S6}}
\end{figure*} 

In order to compare the partition obtained with the different method we have plotted in Figure S\ref{S5} and S\ref{S6}, respectively, the contingency tables between the partitions obtained with Infomap and Oslom and Infomap and Louvain for each case study. In these figures, each plot represents a contingency table $C$ in which each element $C_{ij}$ is the number of nodes which belong to the cluster $i$ detected with Infomap and to the cluster $j$ detected with Oslom or Louvain. The matrices have been normalized so that the sum of each column is equal to one. This normalization allows us to study how the nodes belonging to the groups obtained with Oslom or Louvain are distributed among the clusters found with Infomap. First, we can observe that the number of communities detected with Louvain or Oslom is always greater or equal to the ones obtained with Infomap. Indeed, Louvain has detected a similar number of clusters whereas the number of communities detected with Oslom increases with the size of the metropolitan area, from $5$ clusters for Bilbao to $12$ for Madrid. However, it is worth noting that in most of the cases, more than $80$\% of the nodes belonging to the Oslom and Louvain's clusters are gathered in one Infomap cluster. This means that even if the size of the partitions are different, we observe that clusters obtained with Louvain and Oslom are sub-clusters of clusters identified with Infomap.   

\subsection*{Comparison with cadastral data}

In order to validate the results we compared the land use patterns obtained with our algorithm with cadastral data available on the Spanish Cadastral Electronic Site \footnote{~\url{http://www.sedecatastro.gob.es/OVCInicio.aspx}}. The dataset contains information about land use for each cadastral parcel of the metropolitan area of Madrid and Barcelona (about $650,000$ parcels). In particular, we have for each cadastral parcel the net internal area devoted to Residential, Business and Industrial uses. We can use these data to identify the dominant cadastral land use in each grid cell classified as Residential, Business and Industrial uses by the community detection algorithm. To do so we need to define a rule to determine what is the dominant land use in a cell. Intuitively, one would tend to identify the dominant land use in a cell as the land use class with the largest area. However, Residential use is the land use class with the largest area in most of the cell leading to an over-representation of Residential cells in the metropolitan area. To circumvent this limitation we introduce two thresholds $\delta_{Bus}$ and $\delta_{Log}$ to identify Business and Logistics cells with cadastral data. If the fraction of area devoted to Business in a grid cell is higher than $\delta_{Bus}$ then the grid cell is classified as Business. Otherwise, if the fraction of area devoted to Logistics is higher than $\delta_{Log}$ then the grid cell is classified as Industry. Finally, if the fraction of area devoted to Business and Logistics is, respectively, lower than $\delta_{Bus}$ and $\delta_{Log}$ then the grid cell is classified as Residential.

\begin{table*}
	\caption{\sf \textbf{onfusion matrix of the classification for Madrid and Barcelona.} For the Residential, Business and Logistics rows and columns, the value in the $i^{th}$ row and the $j^{th}$ column gives the percentage of grid cells classified as use $i$ by the cadastral classification which are classified as belonging to the class $j$ by the algorithm. The Total is the distribution of the percentage of cells according to the land use type obtained with our algorithm (row) and the cadastral data (column) with the threshold values $\delta_{Bus}=0.2$ and $\delta_{Log}=0.2$.}
	\label{ConfM}
	
		\begin{center}
			\begin{tabular}{|l|c|c|c|c|}
			  \multicolumn{5}{c}{\textbf{Madrid}}\\
				\multicolumn{5}{c}{}\\
				\hline
				\centering \backslashbox{\textbf{Cadastral}}{\textbf{Algorithm}}  & \textbf{Residential} & \textbf{Business} & \textbf{Logistics} & \textbf{Total}\\
				\hline
				\textbf{Residential} & \textbf{71.23} & 21.67 & 7.1 & 49.04 \\ 
				\hline
				\textbf{Business}    & 30.84 & \textbf{62.33} & 6.83 & 39.55 \\   
				\hline
				\textbf{Logistics}    & 22.9 & 32.06 & \textbf{45.04} & 11.41\\ 
				\hline
				\textbf{Total}       & 49.74 & 40.42 & 9.84 & 100\\ 
				\hline
	  	\end{tabular}
	  \end{center}
		
		\begin{center}
			\begin{tabular}{|l|c|c|c|c|}
			  \multicolumn{5}{c}{\textbf{Barcelona}}\\
				\multicolumn{5}{c}{}\\
				\hline
				\centering \backslashbox{\textbf{Cadastral}}{\textbf{Algorithm}}  & \textbf{Residential} & \textbf{Business} & \textbf{Logistics} & \textbf{Total}\\
				\hline
				\textbf{Residential} & \textbf{68} & 26.55 & 5.45 & 47.5 \\ 
				\hline
				\textbf{Business}    & 28.99 & \textbf{52.17} & 18.84 & 35.75 \\    
				\hline
				\textbf{Logistics}    & 13.4 & 32.99 & \textbf{53.61} & 16.75\\ 
				\hline
				\textbf{Total}    & 45.77 & 37.65 & 16.58 & 100\\ 
				\hline
	  	\end{tabular}
	  \end{center}
\end{table*}

\begin{figure}
		\begin{center}
		   \includegraphics[scale=0.35]{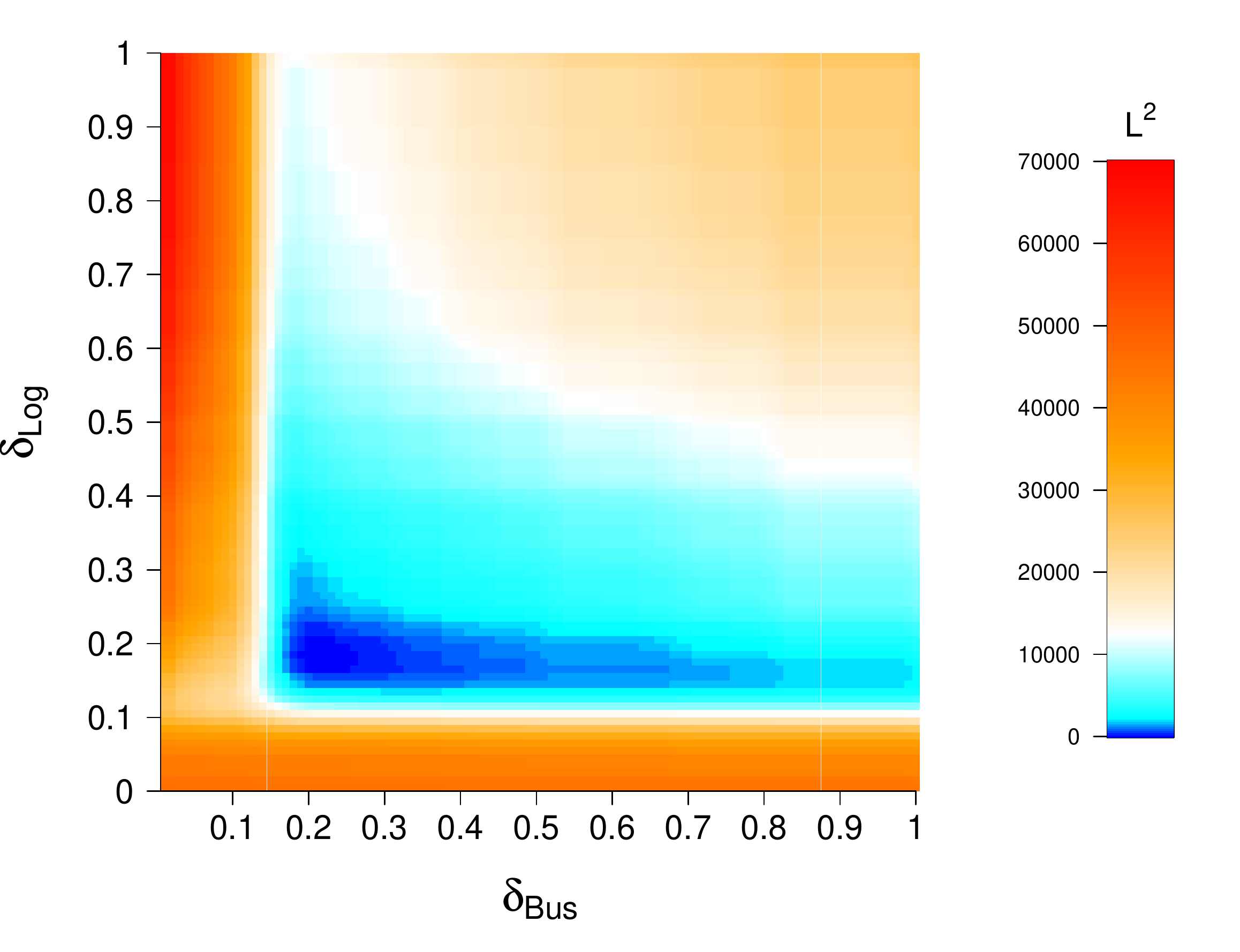}
		\end{center}	
				\caption{\sf $L^2$ distance between the distribution of the fraction of cells according to the land use type (Residential, Business and Logistics) obtained with our algorithm and the cadastral data as a function of $\delta_{Bus}$ and $\delta_{Log}$ for the municipality of Barcelona. \label{S7}}
\end{figure}

Hence, we can adjust the values of these two thresholds in order to obtain a distribution of the fraction of cells according to the land use type similar to the one obtained with our algorithm. To this end we have calibrated these parameters by minimizing the $L^2$ distance between the distribution of the fraction of cells according to the land use type obtained with the cadastral data and the one obtained with our algorithm for the municipality of Barcelona which represents 20\% of the metropolitan area of Barcelona. In Figure S\ref{S7}, we can observe that the minimum is reached for $\delta_{Bus}=0.2$ and $\delta_{Log}=0.2$. Now we can use these values to identify the dominant cadastral land use in each grid cell of the metropolitan area of Barcelona and Madrid.

We find a percentage of correct predictions equal to 65\% for Madrid and 60\% for Barcelona which is consistent with values obtained in other studies, $54\%$ in \cite{Toole2012} and $58\%$ in \cite{Pei2013}. Furthermore, for both case studies, almost all land use types have a percentage of correct predictions higher than $50\%$ (Table S\ref{ConfM}).

\begin{figure*}[t]
		\begin{center}
		   \includegraphics[width=\linewidth]{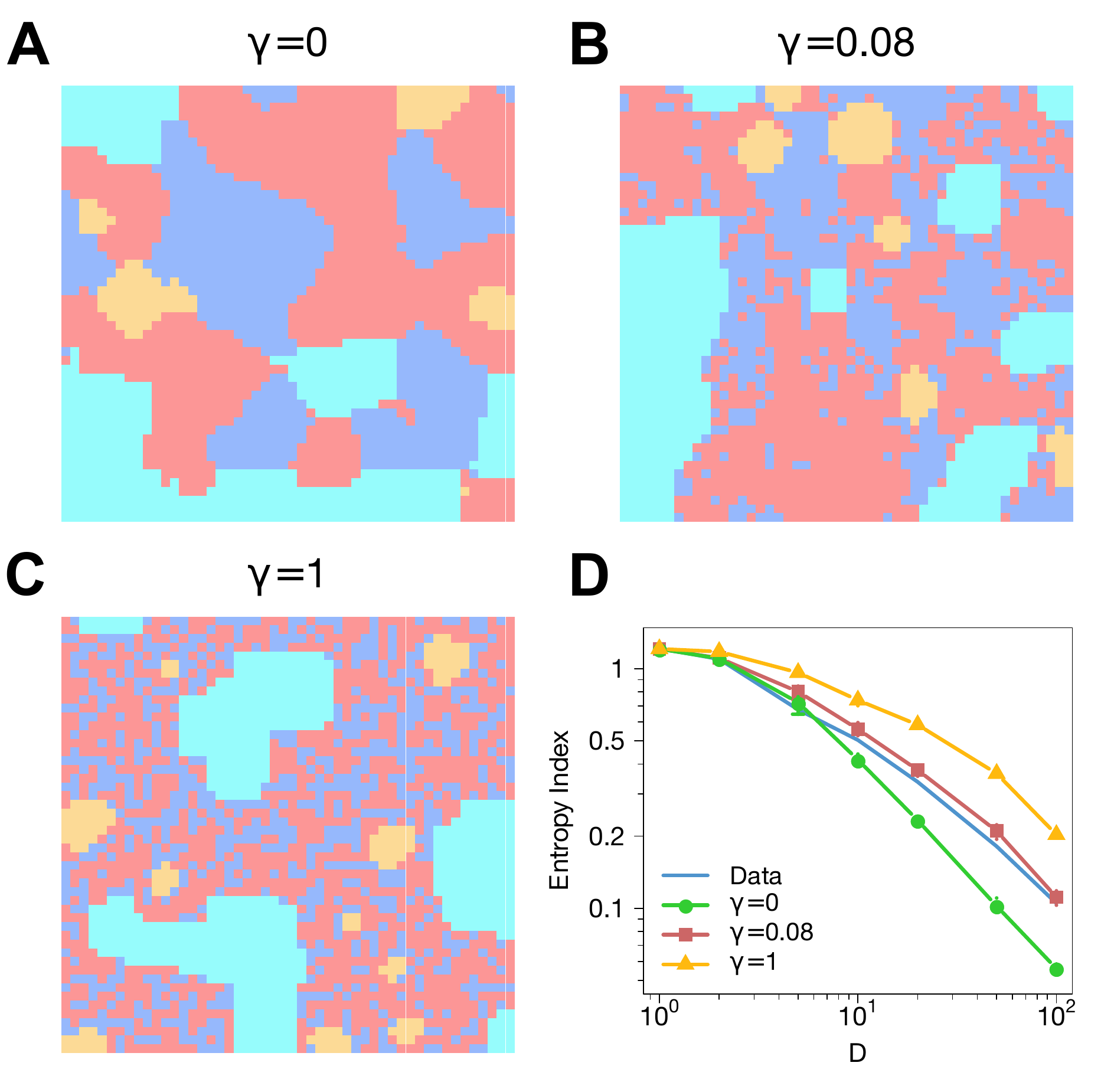}
		\end{center}	
				\caption{\sf Results obtained with different values of $\gamma$ ($\gamma = 0$, $\gamma = 0.8$ and $\gamma = 1$) and $T=500,000$. The 2D lattice used to represent the urban space is composed of $50 \times 50 =2,500$ cells. The model seems to converge after $300,000$ iterations but to ensure the convergence all the results shown in the paper were obtained with $500,000$ iterations.\label{S8}}
\end{figure*}

\subsection*{Calibration of $\gamma$}	

The value of $\gamma$ was calibrated in order to reproduce the evolution of the entropy index as a function of the number of divisions by side obtained with the data (red line in Figure 4 in the main text). We chose the value of $\gamma$ minimizing the Euclidean distance between the observed values and the average values obtained with the model with $100$ replications. The best results have been obtained with the value $\gamma=0.8$ (Figure S\ref{S8}).

\clearpage
\begin{figure*}
		\begin{center}
		   \includegraphics[scale=0.15]{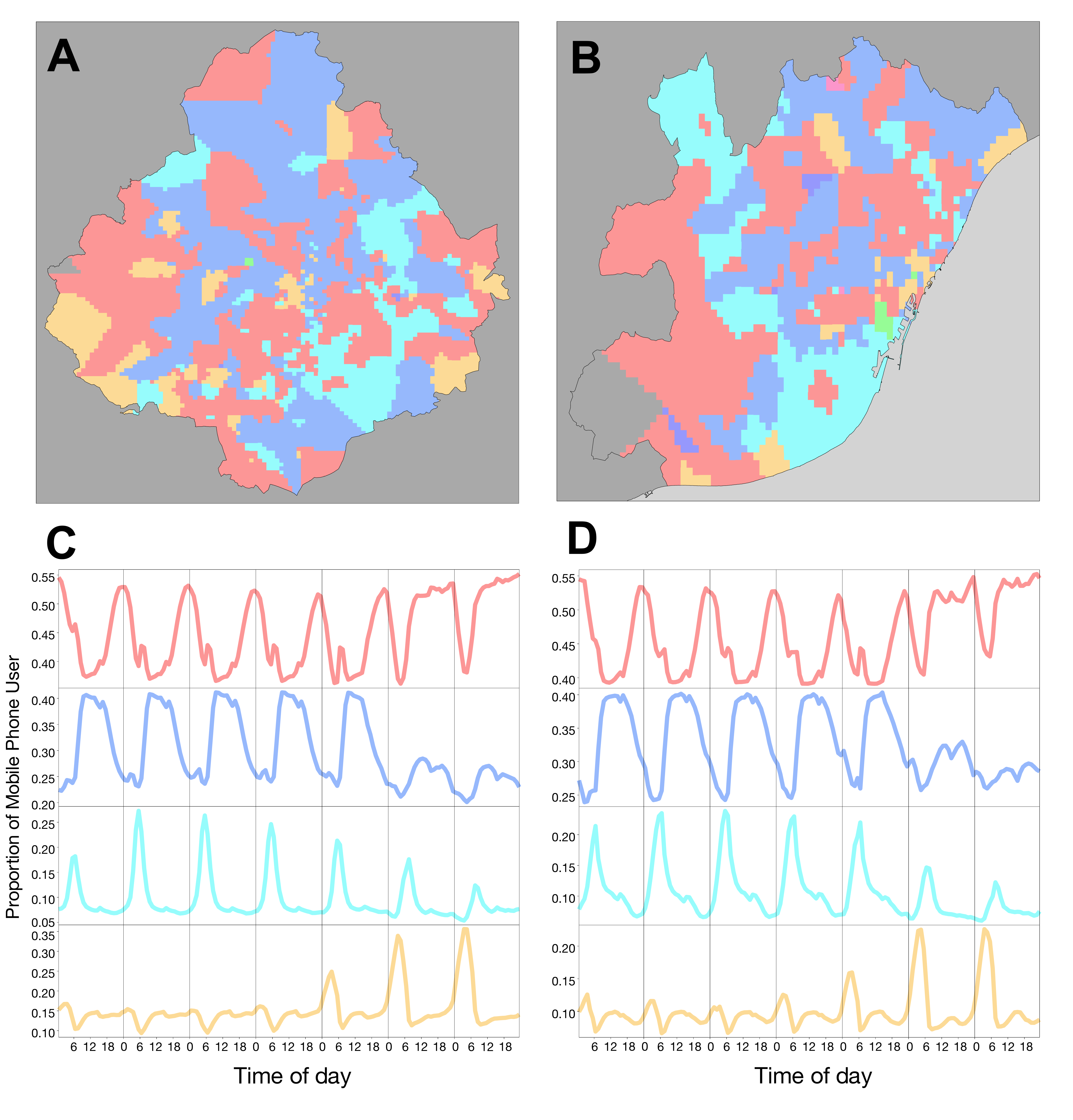}
		\end{center}	
				\caption{\sf (A-B) Geographical location of the clusters for Madrid and Barcelona. (C-D) Temporal patterns associated with the four clusters for both metropolitan areas. In red, Residential cluster; In blue, Business; In cyan, Logistics; And in orange, Nightlife. \label{S9}}
\end{figure*}

\begin{figure*}
		\begin{center}
		   \includegraphics[scale=0.12]{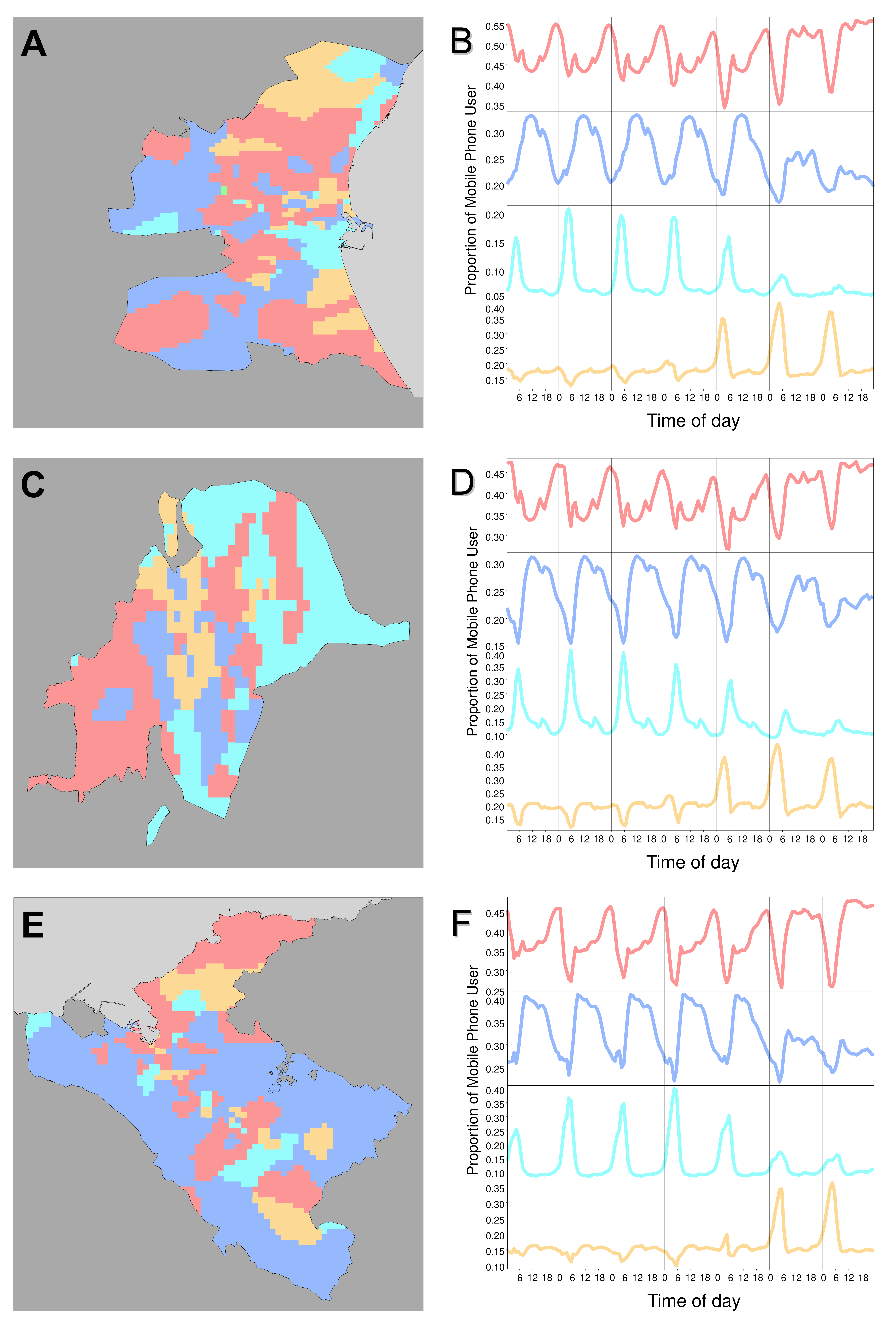}
		\end{center}	
				\caption{\sf (A), (C) and (E) Geographical representation of the communities for Valencia (A), Sevilla (C) and Bilbao (E). (B), (D) and (F) Temporal patterns associated with the communities for the metropolitan area of Valencia (B), Sevilla (D) and Bilbao (F). In red, the Residential community; In blue, the Business community; In cyan, the Logistics/Industry community; In orange, the Nightlife community. \label{S10}}
\end{figure*}

\end{document}